\documentclass[12pt]{article}
\usepackage{epsf}

\usepackage{epsfig,graphics}
\usepackage {graphicx}
\usepackage {epsfig}
\usepackage {subfigure}
\usepackage {tabularx} 
\usepackage{rotate}	
\usepackage{slashed}

\usepackage{amsmath}
\usepackage{amsfonts}
\usepackage{amssymb}
\usepackage{graphicx}
\usepackage{cite}

\usepackage{fancyhdr}
\usepackage{hyperref}

\newcommand{\bmat}{\left(\begin{array}}
\newcommand{\emat}{\end{array}\right)}

\def\yzero{\smash{\hbox{$y\kern-4pt\raise1pt\hbox{${}^\circ$}$}}}

\def\beq{\begin{equation}}
\def\eeq{\end{equation}}
\def\beqa{\begin{eqnarray}}
\def\eeqa{\end{eqnarray}}

\def\-{\hphantom{-}}

\def\s2{\frac{1}{\sqrt2}}

\def\beq{\begin{equation}}
\def\eeq{\end{equation}}
\def\beqa{\begin{eqnarray}}
\def\eeqa{\end{eqnarray}}

\def\IF{\relax{\rm I\kern-.18em F}}
\def\II{\relax{\rm I\kern-.18em I}}
\def\IP{\relax{\rm I\kern-.18em P}}
\def\IC{\relax\hbox{\kern.25em$\inbar\kern-.3em{\rm C}$}}
\def\IR{\relax{\rm I\kern-.18em R}}

\def\Dsl{\,\raise.15ex\hbox{/}\mkern-13.5mu D} 
\def\IZ{Z\kern-.4em  Z}

\catcode`\@=11   
\newdimen\@rotdimen
\newbox\@rotbox  

\def\@vspec#1{\special{ps:#1}}
\def\@rotstart#1{\@vspec{gsave currentpoint currentpoint translate
   #1 neg exch neg exch translate}}
\def\@rotfinish{\@vspec{currentpoint grestore moveto}}
\def\@rotr#1{\@rotdimen=\ht#1\advance\@rotdimen by\dp#1%
   \hbox to\@rotdimen{\hskip\ht#1\vbox to\wd#1{\@rotstart{90 rotate}%
   \box#1\vss}\hss}\@rotfinish}
\def\@rotl#1{\@rotdimen=\ht#1\advance\@rotdimen by\dp#1%
   \hbox to\@rotdimen{\vbox to\wd#1{\vskip\wd#1\@rotstart{270 rotate}%
   \box#1\vss}\hss}\@rotfinish}%
\def\@rotu#1{\@rotdimen=\ht#1\advance\@rotdimen by\dp#1%
   \hbox to\wd#1{\hskip\wd#1\vbox to\@rotdimen{\vskip\@rotdimen
   \@rotstart{-1 dup scale}\box#1\vss}\hss}\@rotfinish}%
\def\@rotf#1{\hbox to\wd#1{\hskip\wd#1\@rotstart{-1 1 scale}%
   \box#1\hss}\@rotfinish}%
\def\rotate{\@ifnextchar[{\@rotate}{\@rotate[l]}}
\def\@rotate[#1]#2{\setbox\@rotbox=\hbox{#2}\@nameuse{@rot#1}\@rotbox}

\catcode`\@=12

\topmargin
-1.5cm
\textwidth
15.5cm
\textheight
23.5cm
\oddsidemargin
0.7cm
\evensidemargin
0.7cm

\begin{document}

\makeatletter
\@addtoreset{equation}{section}
\makeatother
\renewcommand{\theequation}{\thesection.\arabic{equation}}
\pagestyle{empty}
\vspace{-0.2cm}
\rightline{ IFT-UAM/CSIC-16-111}
\vspace{1.2cm}
\begin{center}


\LARGE{ An Axion-induced  SM/MSSM Higgs  Landscape \\
and the Weak Gravity Conjecture\\ [13mm]}

  \large{Alvaro Herr\'aez and Luis E. Ib\'a\~nez \\[6mm]}
\small{
  Departamento de F\'{\i}sica Te\'orica
and Instituto de F\'{\i}sica Te\'orica UAM/CSIC,\\[-0.3em]
Universidad Aut\'onoma de Madrid,
Cantoblanco, 28049 Madrid, Spain 
\\[8mm]}
\small{\bf Abstract} \\[6mm]
\end{center}
\begin{center}
\begin{minipage}[h]{15.22cm}
We construct models in which the SM Higgs mass scans in a landscape.  This is achieved by coupling 
the SM to a monodromy axion field   through Minkowski 3-forms. The Higgs mass scans with steps
given by $\delta m_H^2\simeq \eta \mu f$, where $\mu$ and $f$ are the axion mass and periodicity respectively, and $\eta$ measures the
coupling of the Higgs to the associated 3-form.
 The observed Higgs mass scale could
then be selected on anthropic grounds.  
The monodromy axion may have a mass  $\mu$  in a very wide range depending on the value of $\eta$,
and the axion periodity $f$. For $\eta\simeq 1$ and $f\simeq 10^{10}$ GeV, one has $10^{-3}eV\lesssim \mu \lesssim 
10^3 eV$, but ultralight axions with e.g. $\mu \simeq 10^{-17}$ eV are also possible.
In a different realization we consider  landscape models coupled to the MSSM. 
In the context of SUSY, 4-forms appear as being part of the auxiliary fields
of SUSY multiplets.  The scanning in the 4-forms thus translate into a landscape of
vevs for the $N=1$ auxiliary fields and hence as a landscape for the soft terms.  This could 
provide a rationale for the MSSM fine-tuning suggested by LHC data.
In all these models there are 3-forms coupling
to membranes which induce transitions between different vacua through bubble nucleation.
The {\it Weak Gravity Conjecture}  (WGC) set limits
on the tension of these membranes and implies new physics thresholds well below the Planck scale. More generaly,  we argue that
in the case of string SUSY vacua in which the Goldstino multiplet contains a monodromy axion the WGC 
suggests a lower bound on the SUSY breaking scale $m_{3/2}\gtrsim M_s^2/M_p$.

\end{minipage}
\end{center}
\newpage
\setcounter{page}{1}
\pagestyle{plain}
\renewcommand{\thefootnote}{\arabic{footnote}}
\setcounter{footnote}{0}



\tableofcontents

\section{Introduction}

There are a couple of very bizarre small mass scales in physics. One is the cosmological constant which, if identified with 
dark energy,  is of order $V_0\simeq (10^{-3}eV)^4$, ridiculously small compared to any other scale in the theory. The other is the
Electro-Weak(EW) scale which is of order $m_H\simeq 10^2$ GeV, much smaller than any expected ultraviolet(UV) cut-off. 
Possibly the best solution to the first question was suggested by Weinberg \cite{Weinberg:1988cp}, who pointed out thar if the c.c. $V_0$ {\it scans} in 
a large multiplicity of finely-grained values, galaxy formation requires  $V_0$ to be positive and of order the presently observed values. This is a remarkable 
prediction, since it was pointed out before the existence of dark energy was confirmed.

A natural question is whether an  analogous mechanism could be at work for the Higgs hierarchy problem. The EW scale is tied up to the
mass parameter  $m_H^2$ of the Higgs boson, which is unstable under radiative corrections and would be expected to be of order the cut-off scale 
$m_H^2\simeq \Lambda_{UV}^2$. One way to stabilize the Higgs mass is low energy SUSY. However the observed relatively large Higgs mass
suggests that SUSY, if present, is possibly beyond the reach of LHC or much heavier. So, even though SUSY still remains the
most ellegant solution to the hierarchy problem it makes sense to look for alternative or complementary solutions. 

In the present paper we study  the generation of a {\it landscape} of Higgs mass parameters  $m_H^2$ to address the EW hierarchy problem.  This landscape 
will contain a large number of possible values for $m_H^2$ from large and negative (or positive) to small with $m_H^2$  in the observed phenomenological range.
For the observed value of the EW scale to be one of the possibilities in the landscape, we need  $m_H^2$ to scan with  a fine-grain mass scale
a fraction of the EW mass scale $m_W$. In fact anthropic considerations require  the EW vev not to be far from the measured value 
$v_0= 170$ GeV. Defining    $v=v_0+\delta v$  one finds constraints \cite{Damour:2007uv,Donoghue:2009me,Hall:2014dfa,Meissner:2014pma,Donoghue:2016tjk}
\beq
0.39 \ \leq \ \frac {|v_0+\delta v|}{ v_0} \ \leq \ 1.64   \ 
\label{donaghue}
\eeq
which implies
\beq
\frac {\delta m_H^2}{m_{H_{0}}^2} \ = 2\frac {\delta v}{v_0} \leq  \ 1.2
\eeq
These limits come essentially from the {\it atomic principle}, i.e. imposing that complex and stable nuclei can form.  Note that it requires $\delta v\leq  0.6 v_0$ GeV and hence
practically determine the weak scale to be what experimentally  is.  These constrains may be considered  a necessary but not a sufficient condition 
for an anthropic solution to the hierarchy problem. Indeed, it is well known that 
 the masses of the first generation quarks and leptons would also need to scan in an anthropic setting, see e.g. \cite{Damour:2007uv,Donoghue:2009me,Hall:2014dfa,Meissner:2014pma,Donoghue:2016tjk}. In this paper we will only address the issue of a landscape of Higgs mass parameters which is necessary for an antropic 
 solution to work. Note in this connexion that we we will not try to look  here for
Higgs mass distributions which are peaked around the EW scale.  For an anthropic solution of
the hierarchy problem it is enough to show that there is a landscape of Higgss masses which contains 
the observed Higgs mass, it does not need to be the most likely value. The purpose of this paper is to construct  models
in which indeed the Higgss mass scans and hence completes the above {\it atomic principle} into a possible solution to  the hierarchy problem.
For a discussion of  some phenomenological  scenarios from a field theory landscape  see also  \cite{ArkaniHamed:2005yv}.

\begin{figure}[ht]
\centering
{\includegraphics[width=0.6\textwidth]{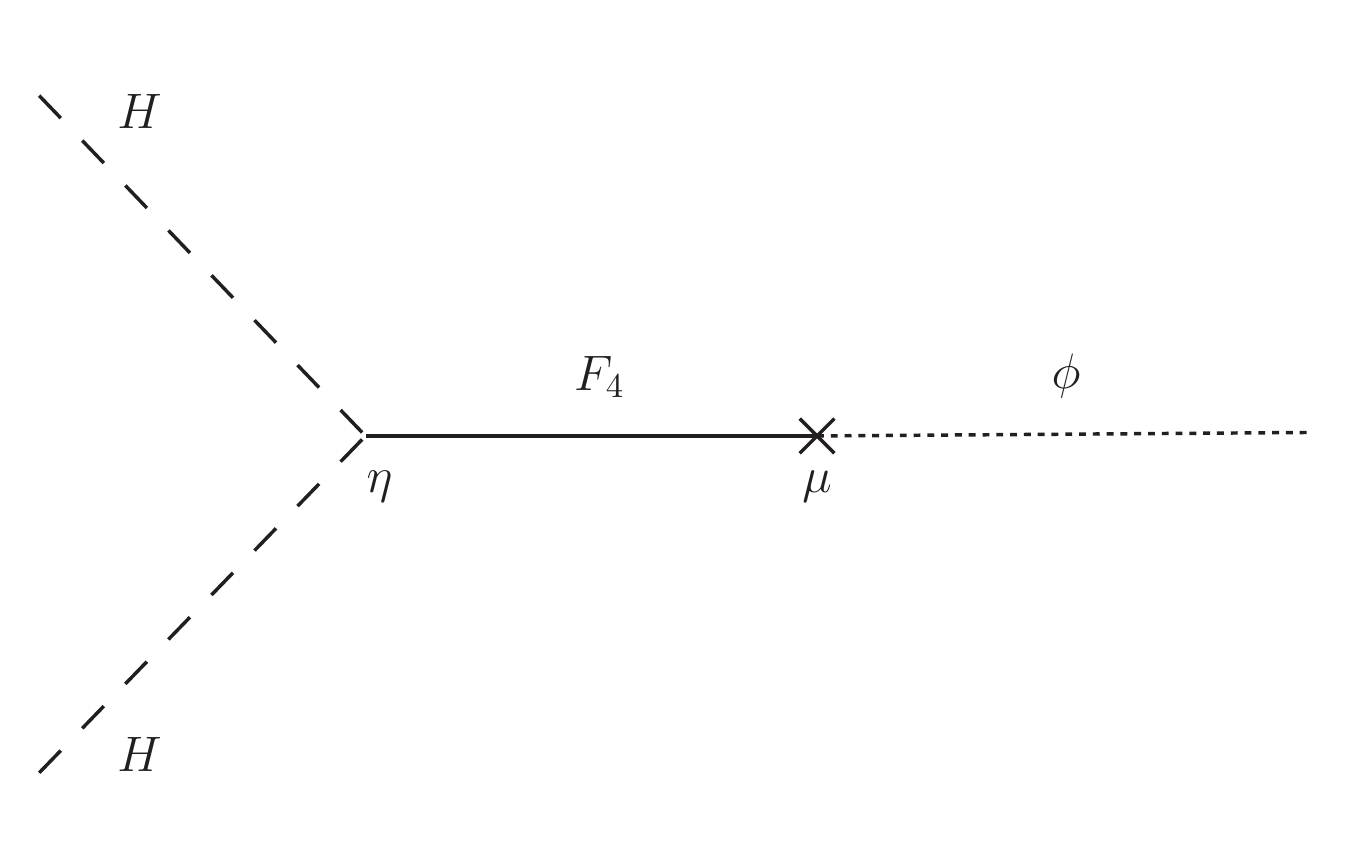}}
\caption{\small{Scheme of the Higgs-axion system. As the axion completes one cycle as $\phi\rightarrow \phi+ f$, the Higgs mass$^2$
changes by $\delta(m_H^2)\simeq \eta \mu f$ .}}
\label{scanning}
\end{figure}

We consider two classes of models, non-SUSY and SUSY, with some important differences between them. In both cases the important ingredient is the existence of
Minkowski 3-forms $C_3$  \footnote{For pioneer works involving 3-forms see e.g.
\cite{duff1,hawking,duff2,wu,duncan,BT,BP,MR}. For more recent applications see 
\cite{Dvali1,KS,KLS,imuv,biv}. Four-form-Higgs couplings were considered by Dvali in\cite{Dvali1}.}, which couple to the Higgs sector via their field strength $F_4$. These 4-forms in turn couple to an axion-like field $\phi$ in such a way that a
built-in shift symmetry  under $\phi\rightarrow \phi +f$ is respected. The 4-forms are quantized in units of $\mu f$, with $\mu$ the axion mass. In the simplest  non-SUSY example
the 4-form couples to the Higgs in a renormalizable way,  with a term $\eta F_4 |H|^2$ (see fig.(\ref{scanning})). This coupling makes the Higgs mass to scan in a landscape of values in steps
given by $\eta \mu f\lesssim m_W^2$. The mass of this axion-like field (or {\it Hierarxion}) is hence  of order $\mu \lesssim  m_W^2/f$, which is tipically very small.

In the SUSY case, the 4-forms are part of the auxiliary field system of the $N=1$ multiplets. The coupling of the Higgs system to the 4-forms
appear as in standard gravity mediation, so the Higgs fields get mass$^2$ of order $F_4^2/M_p^2$. This suggests to identify the vev of $F_4$ with
an intermediate scale, $F_4\simeq (10^{10})^2$ GeV$^2$, so that one obtains Higgs masses of order the EW scale. Within string theory the 3-forms associated to
these 4-forms couple to membranes whose tension would be typically of order the string scale, i.e. $(F_4^{3/2})\simeq M_s^3$.  Thus the string scale will typically
be of order the intermediate scale, $M_s\simeq 10^{10}$ GeV.  Again,  the fact that the 
auxiliary fields related to the 4-forms are quantized, makes the Higgs mass and in general all soft terms to scan in a landscape of values around the EW scale.
In particular, the possibility exists that soft terms could be relatively large compared to LHC scales, say 3-5 TeV and still having  correct fine-tuned EW breaking 
by a fortuitous cancellation of different 4-form contributions in the landscape.

The 3-forms couple to membranes which may nucleate transitions in these landscapes of vacua. 
The Weak Gravity Conjecture (WGC)\cite{WGC} \footnote{ See  e.g. \cite{WGC1,imuv,WGC2,Garcia-Valdecasas:2016voz} for recent WGC papers.}
strongly constraints the  tension of the membranes involved, as well  as the mass of the
axion. In the non-SUSY examples  the membrane tension is bounded above by $T \lesssim \eta^{-1} (10^8)^3$ GeV$^3$, with $\eta$ the
4-form-Higgs coupling. The corresponding axion is bounded below as $\mu_{axion} \gtrsim T/2\pi f M_p$, and impossing further stability against nucleation 
one has $\mu _{axion}\gtrsim  10^{-3}$ eV $(10^{10}GeV/f)(m/10^{10}GeV)^{3/2}$, with $m$ the UV Higgs cut-off. In the SUSY examples 
the WGC suggests that $m_{3/2} \gtrsim  T/fM_p$. Within string theory one expects $T^{1/3}\simeq f\simeq M_s$, with $M_s$ the string scale,
so that one gets $m_{3/2} \gtrsim  M_s^2/M_p$.  This implies that if we want to have SUSY breaking soft terms of order the EW scale, the string scale should
be of order the intermediate scale $M_s\simeq \sqrt{M_WM_p}\simeq 10^{10}$ GeV or below. This is a very strong constraint, since the 
typical scenario with $M_s\simeq 10^{16}$ GeV would be ruled out in this context.

More generally,  we argue  that in  certain classes of string compactifications  in which SUSY is broken by fluxes,  and the Goldstino multiplet
contains a monodromy axion, the Weak Gravity Conjecture suggests a lower bound 
on the SUSY breaking scale $m_{3/2}$
\beq 
m_{3/2} \ \geq \ \frac {M_s^2}{M_p} \ .
\eeq
This bound has a number of loopholes, some of which are discussed in the text. Still, if true, it would have important phenomenological implications. For example,
having low energy SUSY at 1 TeV would require a string scale at $10^{10}$ GeV.

The structure of the rest of this paper is as follows. In the next section we review a few facts about Minkowski 3-forms and their interaction with
axions. In section 3 we construct a minimal (non-SUSY) model in which a Higgs mass landscape is generated in terms of quantized shifts 
of an axion. We also study the instability of the model against buble nucleation and constraints on the axion mass and scale of new physics 
from the WGC. Limits on the mass of these axions are given.  In section four we address the construction of $N=1$ SUSY models with a
Higgs mass landscape. We discuss the mentioned lower bound on the SUSY breaking scale from the WGC in section five and  leave the last section for some general comments
and conclussions.

\section{Axions and 3-forms}

Before presenting the model let us briefly review a few facts about these Minkowski 3-forms  (see e.g. \cite{BT,BP,MR,Dvali1,KS,KLS,msu,domainwalls,Dudas}). The action 
for a 3-form $C_{\nu \rho \sigma}$ is given by
\beq
{\cal L}\ =\  -  \frac {1}{2}  F_{\mu\nu\rho\sigma}F^{\mu \nu \rho\sigma} \ +\  {\cal L}_{bound} \ 
\eeq
where $F=dC$ is the field-strength 4-form and  ${\cal L}_{bound}$ includes some boundary terms which, although necessary to get 
the right field equations (see e.g. the discussion in  \cite{KS,Dudas}),  will not be relevant in our discussion.  The equations of motion 
imply that the 4-form is a constant tensor in Minkowski,
\beq
F_{\mu\nu\rho\sigma}\ =\ f_0\epsilon_{\mu\nu\rho\sigma}
\eeq
where $f_0$ is a real constant of mass dimension two. Note that $f_0$ behaves  as a constant electric 4-form field permeating
the whole Minkoski space and contributing (positively) to the vacuum energy in a way proportional to $f_0^2$. 
 We see that  a 3-form has no propagating degrees of freedom. Still it may have interesting dynamics. In particular, 3-forms
naturally couple to the worldvolume of membranes (or domain walls) through
\beq
S_{mem}\ =\ q\int_{D_3} d^3\xi \epsilon^{abc}\ C_{\mu\nu\rho}\left(
\frac {\partial X^\mu}{\partial\xi^a}\frac {\partial X^\nu}{\partial\xi^b}\frac {\partial X^\rho}{\partial\xi^c}\right) \ ,
\eeq
where the membrane charge $q$ has dimensions of mass$^2$ and $D_3$ is the membrane world volume. Due to this coupling, regions of space separated 
by  membranes  change their 4-form background $f_0$ by
\beq
f_0\ \longrightarrow \ f_0 \ +\ nq  \ \ ,\ \ n\in {\bf Z} \ .
\label{cuantiza}
\eeq
So membrane nucleation yield  changes in the value of the 4-form background which are quantized in units of the
membrane charge.  At this level the value of the charge $q$ is undetermined. However  there is an interesting way in which
the value of $q$ turns out to be constrained.   This happens if the 3-form is coupled to an axion-like 
scalar giving it a mass.  Let us introduce an axion-like field $\phi$, i.e., a pseudoscalar with a 
discrete  shift symmetry under
\beq
\phi \ \longrightarrow \ \phi \ + \ m f   \ \ ,\ \ m\in {\bf Z}     \ ,
\eeq
with $f$ the axion periodicity.
Let us consider the addition to the action of a direct coupling of the axion to the 4-form
\beq
{\cal L}\ =\ -\frac12(\partial_{\mu} \phi)^2-\frac12 F^2\  +  \ \mu \phi F \ .
\eeq
Using the equations of motion for $F$ one obtains a scalar potential
\beq
V\ =\ \frac {1}{2} |f_0\ + \ \mu \phi|^2   \ ,
\eeq
where we have allowed for a 4-form vev $f_0$.
Note that, even though now the axion has mass  $\mu$, the axion shift symmetry is respected if the
4-form also transforms apropriately
\beq
\phi \ \rightarrow \ \phi \ + \ nf\ \ ,\ \  f_0\ \rightarrow \ f_0 \ -\ n\mu f  \ .
\label{shifts}
\eeq
Comparing eqs. (\ref{cuantiza}) and (\ref{shifts}) one obtains the consistency condition
for the charge $q$ of membranes coupling to this axion system
\beq
nq\ =\ \mu f  \ ,\ n\in{\bf Z}  \ .
\label{generalquanta}
\eeq
This equation relates the otherwise undetermined membrane charge $q$ to the axion parameter product $\mu f$.
This constraint will be interesting below, when we construct a specific model couple to the Higgs. In what follows we will assume
take  $|q|=\mu f $ as the natural value for the 4-form quanta and briefly discuss the more general case below.
This process in which the axion gets mass may be understood as a generalized Higgs mechanism in which  the 
2-form $B_{\rho \sigma }$ dual to the axion field is swallowed and gains a mass $\mu$. Indeed after this duality the 
mass term becomes
\beq
-\ \frac {\mu ^2}{2}|C_3\ -\ dB_2|^2  \ ,
\eeq
which indeed realizes a generalized Higgs mechanism. This dual formulation displays in a very explicit way the gauge
origin of the shift symmetry which is at the root of the stability of the axion potential under higher order corrections.

Let us close this section by noting that 
Minkowski four-forms  appear naturally in string theory upon reduction to four dimensions of higher dimensional
RR and NS antisymmetric fields, see e.g.   \cite{BP,MR,KS,msu,domainwalls,biv}.

\section{ A Higgs landscape  from axion monodromy. A minimal model}

Let us now couple this axion/3-form system to the SM Higgs field $H$. For reasons to be obvious later the minimal model one
can build involves two 4-forms $F_a$ and $F_h$ (esentially what happens is that with a single 4-form the axion vev is fixed
in terms of the Higgss vev, but there is no scanning effect). By definition  $F_h$ the linear combination of the two 4-forms 
which couples to the Higgs through a dim=4 operator. The relevant piece of the action is then
\footnote{ Note that this is a two 4-form generalization of the relaxion model constructed in section (3.2) of ref.\cite{imuv}. However 
here we are not considering a relaxion type of model \cite{relaxion}
and cosmology plays no crucial role in this landscape construction.}
\beq
{\cal L}  \ =\ -\frac {1}{2}(F_a)^2 \ -\frac {1}{2}(F_h)^2\ +\ \phi (\mu_a F_a+\ \mu_h F_h)\ +\ \eta F_h|H|^2 \ .
\eeq
Here $\eta$ is an adimensional coupling constant.
Using the equations of motion for the 4-forms one finds the potential 
\beq
V\ =\ \frac {1}{2} |f_0^a\ +\ \mu \phi|^2 \ +\ \frac {1}{2} |f_0^h\ +\ \mu^h \phi \ +\ \eta \sigma^2|^2  \ ,
\eeq
where we have set the Higgs to its physical neutral component $|H|^2 =\sigma^2$. 
Note that this scalar potential is invariant under the axion shift symmetry
\beq
\phi \rightarrow \phi\ +\  nf \ ;\ f_a^a\rightarrow f_0^a\ - \   \mu nf  \  ;\   f_0^h\ -\   \mu^h nf  \ , \ n\in {\bf Z} .
\eeq
The membranes coupling to these 3-forms will have charges $q_a,q_h$  related to the axion parameters as
\beq
q_a\ =\ \mu f    \ ,\   q_h\ =\  \mu^h f    \  .
\label{quanta}
\eeq
The above shift symmetry guarantees that the mass parameters $\mu, \mu_h$ are stable under loop corrections,
the form of the axion dependent potential above will remain even after these corrections. 
On the other hand the Higgs field 
couples to the full SM through gauge and Yukawa interactions which will induce masses and quartic coupling corrections. 
Thus the scalar potential will have really the form
\beq
V\ =\ \frac {1}{2} |f_0^a\ +\ \mu \phi|^2 \ +\ \frac {1}{2} |f_0^h\ +\ \mu^h \phi \ +\ \eta \sigma^2|^2  \   - \  m^2\sigma^2 \ +\ \lambda \sigma^4
\eeq
once corrections are taken  into account. Here $m^2$ will typically be of order the UV scale, since the Higgss mass is unprotected.
The minimization conditions require
\beqa
\partial V/\partial \sigma\ &=&\ \sigma[ 2\eta (f_0^h+\mu^h\phi+\eta \sigma^2)\ +\ 4\lambda \sigma^2\ -\ 2m^2 ]\ =\ 0 \\
\partial V/\partial \phi\ &=&  \mu(f_0^a+\mu \phi) \ +\  \mu^h(f_0^h+\mu^h\phi+\eta \sigma^2)\ =\ 0 \ .
\eeqa
One then finds
\beq
\phi\ =\ -  \ \frac {f_0^a\mu\ +\ f_0^h\mu^h\ +\ \eta \mu^h\ \sigma^2}{\mu^2\ +\ (\mu^h)^2} \ ,
\eeq
with the Higgs vev given by
\beq 
\sigma^2 \ =\ \frac {m^2 - \eta (cos^2\theta  f_0^h\ -sin\theta cos\theta f_0^a)\  }
{2\lambda\ +\ \eta^2 cos^2\theta}
\eeq 
where 
\beq
sin^2\theta  \ =\ \frac { (\mu^h)^2}{\mu^2 + (\mu^h)^2} \ ,\   cos^2\theta  \ =\ \frac { (\mu )^2}{\mu^2 + (\mu^h)^2}  \ .
\eeq
The mass$^2$ matrix of both scalars
has the form
 \beq
M^2 \ =\ 
\left(
\begin{array}{cc}
 { \mu^2+\mu_h^2} &   2 \eta \mu_h \sigma_{min}\\   
  { 2\eta \mu_h \sigma _{min}} & 2{\cal M}^2_{\sigma \sigma}\\
\end{array}
\right)
\label{matrizmasas}
\eeq
with 
\beq
{\cal M}_{\sigma \sigma}^2 \ =  \ 
\frac {4\lambda +2\eta^2}{2\lambda+\eta^2cos^2\theta} \left( m^2\ -\ \eta(f^h cos^2\theta- f^asin\theta cos\theta)\right) \ .
\label{masahiggs}
\eeq
The Higgs vev  at the minimum can also be written in terms of ${\cal M}_{\sigma \sigma}$,
evaluated at the minimum
\beq 
\sigma^2 \ =\  \frac {M_{\sigma \sigma} ^2} {4\lambda+2\eta^2 }  \ .
\label{esperado}
\eeq
Looking at eq.(\ref{masahiggs}) and (\ref{esperado}) we see that  the Higgs  vev scans in a landscape as we vary the 
4-form vevs $f^h,f^a$.  There are always potentials in which  the Higgs vev obeys eq.(\ref{donaghue})  as long as the step of the 4-forms
$q_h, q_a$ are of order  the observed $m_H^2$ or smaller.  In particular if we change the 4 forms by an
amount
\beq
f^h\ \rightarrow f^h\ +\ m_hq_h   \ ,\   f^a\ \rightarrow f^a\ +\ m_aq_a \ \ ,\ \ m_h,m_a\in {\bf Z}
\eeq
then the Higgs vev changes by the amount
\beq
\delta (\sigma ^2) \ =\   \mu f  \frac {  \eta sin\theta cos\theta}{(2\lambda+\eta^2cos^2\theta)} (m_a\ -\ m_h) \  ,
\eeq
or, alternatively, in terms of the 4-form quanta via eq.(\ref{quanta})
\beq
\delta (\sigma ^2) \ =\   \frac {  \eta \mu f }{(2\lambda+\eta^2cos^2\theta)} \frac {q_aq_h}{q_a^2+q_h^2} (m_a\ -\ m_h) \  .
\eeq
Note that if there is no coupling of the Higgs to the axion  ($\eta=0$) there is obviously no possibility of fine-tuning.
Also the two 4-forms are required to couple to the axion so that both $q_a,q_h\not =0$. Assuming both masses $\mu,\mu_h$ 
to be of the same order (in order not to introduce further hierarchies), which also implies 4-form quanta of the same order one can obtain 
a fine-tuning as small as required by imposing
\beq
\delta (\sigma ^2) \ \simeq \ \eta  \mu f \  = \ \eta q_a \ \leq \ m_{H}^2 \ .
\eeq
So the fine-tuning is directly connected to the 4-form quanta $q_a,q_h$ and to the strength of the coupling of the 4-form to the Higgs.
We thus have a large family of SM vacua with different Higgs masses, including a number of them consistent with what is
observed.   Note however that the value of the 4-form values $f_{a,h}$ themselves are very large, of order the Higgs cut-off mass $m^2$,
whereas the membrane charges $q_a,q_h$ are of order the EW scale. This is unlike the SUSY scenario discussed below, in which both are
typically of the same order.

\subsection{Stability and the Weak Gravity Conjecture}

Given the large multiplicity of  Higgs vacua, an interesting question is the stability of these against membrane nucleation.  If these vacua where very short-lived, the
solution to the hierarchy problem would be gone. We can make an estimation 
using the Coleman-De Lucia computation \cite{coleman}  of the transition rate in the thin wall approximation.
The rate is proportional to
\beq 
P\ \simeq \ e^{-B} \ ,\   B\ =\  \frac {27\pi^2 T^4}{2(\Delta V)^3}
\eeq
where $T$ is the tension of the bubbles (membranes) which can nucleate. We can estimate $\Delta V$, which is the change in the
vacuum energy induced by  a change $f_0\rightarrow f_0+q$ in one of the 4-forms, as 
\beq
\Delta V \ \simeq \ qf_0 \ \simeq  \ \frac {q}{\eta} m^2  \ ,
\eeq
where $m^2$ is of order the Higgs cut-off scale, since the 4-form vevs have to cancel a large quadratically divergent 
Higgs mass. On the other hand  actually we do not  know what the tension of the membranes $T$ is.   In any event, from $B>1$ , in order to 
have a supresed rate, assuming that for fine-tuning one also requires $\eta q\simeq m_H^2$,  the tension will be bounded below by
\beq
T \ \gtrsim \ 0.3 \times \eta ^{-3/2}  (m_H^2m^2)^{3/4} .
\label{consnucl}
\eeq
Note that if the associated membranes are fundamental, like e.g. D2-branes in String Theory, a tension $T=(M)^3$ implies the
existence of a new physics scale $M$, like the string scale in the case of String Theory, in which $M=(\alpha')^{-3/2}/g_s$. 
So the above arguments would give a lower bound on such a scale, depending on the size of the Higgs-4-form coupling.
Thus  from stability against nucleation one gets
\beq
M^2 \ \gtrsim \ 0.5\ \eta^{-1} \ m_Hm \ .
\eeq
where $m$ is the UV cut-off of the Higgs mass. For e.g. $\eta\simeq 1$ one has $M \gtrsim  \sqrt{m_Hm} \simeq 10^9$ GeV for
$m\simeq 10^{16}$ GeV. 

To gain further  insight into the mass scales involved we can try to impose further consistency conditions.
In particular  it has been argued that the {\it Weak Gravity Conjecture} extended to 3-forms give us an upper bound on the tension $T$ of membranes coupling to 3-forms. One
has \cite{imuv}
\beq 
T\ \leq \ 2\pi q M_p \ ,
\eeq
where in our case the membrane charge is given by $q=\mu f$, so that we get
\beq
T\ \leq \ 2\pi \mu f M_p \ \leq \ 2\pi\frac { m_H^2}{\eta} M_p \ \simeq \frac {1}{\eta} \ ( {10^{8}}\ GeV)^3  \ ,
\label{conswgc}
\eeq
where again we are assuming here $\eta \mu f\simeq m_H^2$.
Note that by making the coupling small, the tension of the membranes can be made large. For example one
could have  $T\simeq (10^{16}\ GeV)^3$ if $\eta \simeq 10^{-24}$. However if e.g. $\eta\simeq 1$ a threshold of new physics
should appear around or below $10^8$ GeV.

Combining  equations (\ref{consnucl}) and (\ref{conswgc}) one obtains an upper bound on the UV cut-off
coming from imposing supresed nucleation and the WGC constraint given by
\beq
m\ \lesssim \   7.6 \times \eta ^{1/3} (m_HM_p^2)^{1/3}\ \simeq \     {\eta^{1/3}} 10^{14}\ GeV \ .
\label{consUV}
\eeq
Then in this scheme  scalar cut-offs $m$ as large as $10^{14}$ GeV can be fine-tuned in a manner consistent with both
the weak gravity conjecture and stability against nucleation. However this scale $m$ is reduced if the coupling $\eta$ is reduced.

Let us make a couple of comments  about possible slight modifications to  the above results.

1) We have imposed a very conservative  upper bound  on the value of the Higgs mass fine tuning, $\delta m_H^2/m_H^2 \leq 1$.
We can equally consider  a  more finely grained fine-tuning
\beq
\delta m_H^2 \ \leq \ \xi \ m_H^2 
\eeq
with $\xi$ as small as we wish. Then the scales and bounds estimated in eqs(\ref{consnucl}), (\ref{conswgc}) and (\ref{consUV})  remain
applicable replacing in those equations $m_H^2\rightarrow \xi m_H^2$.  In particular the lower bound on the membrane tension
from  stability becomes weaker  whereas the  upper bounds on the tension coming from the WGC becomes stronger, and so happens with 
the UV scale $m$. The dependence on $\xi$ is however weak, due to the $1/3$ power.

2)  In the above estimations  we consider the  quantization constraints $q_a=\mu f$, $q_h=\mu_h f$. One can equally consider the more
general case in which the membrane quanta is an integer fraction of the axion shift, as in eq.(\ref{generalquanta}). All the results above still apply 
replacing $q_a\rightarrow n_aq_a$, $q_h\rightarrow n_hq_h$, with $n_a,n_h\in {\bf Z}$.

Note that one can also obtain a finer tuning  (at fixed $\mu f)$) by reducing the value of the coupling $\eta$ and playing around with the
integers $n_a,n_h$ just mentioned.

\subsection{The {\it Hierarxion} }

One interesting feature of this approach is that there is a new particle,  we may call it the {\it Hierarxion},  which could perhaps have testable properties 
depending on the masses $\mu$, periodicity $f$ and the possible presence of additional couplings to other SM fields beyond the Higgs like e.g. photons.
Concerning the mass of the axion we approximately have
\beq
m_{axion} ^2\ =\  \mu^2+\mu_h^2\ =\ \frac {q_a^2+q_h^2}{f^2}\ \lesssim \ \frac {m_H^4}{\eta^2 f^2}
\eeq
where the latter inequality comes from the fine-tuning condition, assuming  $\xi\simeq 1$. There is also a lower bound on the axion mass if one applies
the WGC argument, since if the quanta $q_{a,h}$ are too small, the interaction of the 3-form with the membranes would be weaker than
the gravitation of the latter, i.e.
\beq
m_{axion} \ \gtrsim  \ \frac {T}{2\pi fM_p} \ \simeq \ \frac {M^3}{2\pi fM_p} \ .
\eeq
Combining it with the stability constraint  $M^2\geq (0.5)\eta^{-1}mm_H$ one has a lower bound
\beq
m_{axion}\ \gtrsim \ \frac {0.3}{2\pi} \eta^{-3/2} \frac {(m_Hm)^{3/2}}{fM_p}
\eeq
We thus see that there is a wide range of possible axion masses.  Depending on the values of the axion periodicity $f$ and the Higgs mass 
UV cut-off $m$ one has 
\beq
4.7\ \eta^{-3/2}\ 10^{-3} eV  \left( \frac {10^{10}GeV}{f}\right) \left(\frac {m}{10^{10}GeV}\right)^{3/2} \ \lesssim  m_{axion} \ \lesssim  
\ \eta ^{-1} \ 10^{3} eV \ \left(\frac {10^{10}GeV}{f}\right)
\label{rango}
\eeq
where we have highlighted the values  for $f\simeq m\simeq 10^{10}$ GeV. Note that as long as the constraint (\ref{consUV}) is fulfilled,
both upper and lower limits are consistent.  As we can see,  a very wide range of values of the hierarxion mass are consistent with
the generation of a SM landscape. For $\eta\simeq 1$ and $f\simeq m\simeq 10^{10}$ GeV  one has $10^{-3}eV\lesssim  m_{axion}\lesssim 10^3 eV$,
but much lighter axions are also possible. Thus for $f\simeq M_p$, $\eta\simeq 1$  and $m\simeq 10^6$ GeV one can have  ultralight axions
with $m_{axion}\simeq 10^{-17}$ eV.
On the other extreme, e.g., if the Hierarxion does not couple directly to gauge bosons and $f\simeq 10^2$ GeV,  the hieraxion could
be as heavy as  hundreds of GeV.  In this case it could mix with the ordinary Higgs.  Furthermore, if the hieraxion-Higgs coupling $\eta$ 
is small, the Hierarxion may have even larger masses.  For example, if $\eta \simeq 10^{-16}$, one could have a Hierarxion mass as large as $10^{10}$ GeV.

This {\it Hierarxion} can give rise to interesting phenomenology which will obviously depend on the values of  the mass and $f$, and also on the 
existence of additional couplings to the SM beyond the necessary coupling to the Higgs.  Here we limit ourselves
to a preliminary discussion and leave a detail discussion for forthcoming work. In particular a light  {\it Hierarxion } could be a dark
matter component.  CMB Planck results  already constrain in an important manner the contribution to dark matter from ultralight axions.
In the region $10^{-32}eV\leq m_{axion}\leq 10^{-26}eV$ the axion contribution to dark matter is less than a few per cent (see e.g.\cite{Marsh:2015xka} and
references therein). On the other hand for axion masses above  $10^{-23}$ eV  an ultralight  { Hierarxion} could constitute most of dark matter.

The {\it Hierarxion} needs not couple to gluons or photons, but if it does, it could perhaps be identified with the QCD axion. 
However the axion potential discussed above can overwhelm the  standard non-perturbative QCD axion potential  and spoil the solution 
to the strong CP problem and render $\theta\simeq 1$.  To avoid that,  one imposes  the constraint 
\beq 
(qm^2/\eta) \ \lesssim  \ \theta_{QCD}  \Lambda_{QCD}^4
\eeq
where  $q$ stands for  $q_{a,h}$ and the $\theta _{QCD}$ is constrained to be $\theta_{QCD}\leq 10^{-10}$.
This means 
\beq
q\ \lesssim  \  \eta \theta_{QCD} \frac {\Lambda_{QCD}^4}{m^2}  \ ,
\eeq
and the EW scale fine-tuning, which is of order $q$,  would be much finer than just $q\simeq m_H^2$. 
Such small value for $q$ however implies, if the WGC applies, that membranes should have a tension
\beq
T\ \leq \ qM_p \ \lesssim \ \eta  \theta_{QCD} \frac {M_p\Lambda_{QCD}^4}{m^2}  \ \lesssim  \  \eta \left( \frac{ 100\ GeV}{m}\right)^2 \ GeV^3 \ .
\eeq
This tension is typically very small, well below the EW scale, and hence we should have observed
the new physics associated to the membranes.
Another possible objection to such small quanta $q$ is that 
membrane nucleation could destabilize the minima through tunneling, as discussed above. 
It is easy to convince oneself using the equations above that the tunneling rate $B$  would  easily
be   $B> 1$ if the scalar mass cut-off obeys
\beq
m^2 \ \lesssim  \  \eta \theta_{QCD}^{1/4}\Lambda_{QCD}M_p \ ,
\eeq
which  is easily obeyed for $m\leq \eta ^{1/2}10^7$ GeV. 

Very massive Hierarxions are also possible.  Looking to the upper limit in 
equation (\ref{rango}) we see that one can have {\it Hierarxions}  with mass  of order 
\beq
m_{axion}\ \simeq \ f\ \simeq  \ 10^2\  GeV   \ .
\eeq
Such  a {\it Megaxion} could be directly detectable at LHC if it couples  to QCD and  photons, since  it  could lead to di-photon events
at an invariant mass in the region of several hundred GeV \footnote{ Models  with that structure were suggested as arising in string
compactifications with a string scale in the TeV scale \cite{megaxions}. In this case the axion is a Ramond-Ramond closed string pseudoscalar.}.
In this case the axion could have a non-negligible mixing with the SM Higgs which could also lead to constraints in the
axion-Higgs system. Note that in this case, having $f$ of order a few hundred GeV implies that some new physics should
appear not much above one TeV in order to restore  perturbative unitarity, which would be violated by the coupling of the {\it Hierarxion} to
gluons at high energies. 

The Hieraxion may also  be superheavy,  with a mass of order the intermediate scale. This is possible if it is very weakly coupled 
to  the Higgs sector. For example if $\eta\simeq 10^{-16}$ one can have a mass of order $10^{10}$ GeV for $f\simeq 10^{10} $ GeV.
Superheavy masses for the Hierarxion are also natural in the context of supersymmetry.

Let us finally  emphasize that the example above, with a linear coupling to the axion $\eta \phi F_4$ is minimal, but is not the
only possibility. One could also consider e.g. quadratic couplings of the form $F_4^2|H|^2/M_{UV}^2$, with $M_{UV}$ some ultraviolet scale.
This structure appears naturally in the SUSY case which we describe below.

\section{A MSSM landscape}

It is interesting to explore whether analogous landscapes could be constructed within $N=1$  SUSY models like e.g. the MSSM. 
It sounds a  bit redundant to introduce SUSY  in theories in which the hierarchy problem is solved via a landscape of Higgss masses. 
However this may be interesting because of several reasons. For example, there are SUSY models in the literature, like Split SUSY \cite{splitsusy}  or Large Scale SUSY 
\cite{Hall:2009nd}
in which the scale of SUSY breaking is very large, of order $10^5-10^{11}$ GeV and the Higgs mass is small by fine-tuning. For those models 
a landscape of soft terms guaranteeing the possibility of a sufficiently light Higgs would be useful. Furthermore one can also consider this type
of fine-tuning in order to understand or motivate the so called ''little hierarchy problem''.  

 For a SUSY version of a landscape we should start  by asking whether there are SUSY multiplets incorporating 
3-forms of the type discussed above. A hint to that is noticing that the Minkowski 4-forms do not propagate, but rather behave like
auxiliary fields.  So it is natural to think that the Minkowski 4-forms could appear as auxiliary fields of some known SUSY multiplets.
Indeed, there are SUSY chiral multiplets in which the usual complex auxiliary field are totally or partially replaced by 4-forms \cite{3formaux,louis3form,Dudas}.
Still these multiplets have not been much studied in the literature.

 Interestingly enough it has been recently shown  \cite{biv,Carta:2016ynn} that this kind of supergravity and supersymmetry multiplets are those which naturally appear 
in  Type II string compatifications in the presence of fluxes . In string compactifications the geometric moduli and the dilaton come along with axion-like scalar fields.
One can show that the dependence of the effective action on the axions comes always through  Minkowski 4-forms, very much like in the 
non-SUSY example above. In the case of Type IIA and Type IIB orientifolds the effective actions contain 4-forms associated to the moduli and
complex dilaton and the scalar potential dependence of the axions appears as a sum of squared 4-forms.  These 4-forms may be identified as 
auxiliary fields of $N=1$ multiplets.

Note that having 4-forms as auxiliary fields is not purely academic since there are a number of physical differences compared to a standard
$N=1$ sugra auxiliary field. In particular the associated 3-forms couple to membranes, which should be present in the theory. The membranes
can nucleate inducing transitions between vacua with different value for the 4-form.  Furthermore the
4-forms are in general quantized. In particular in string-theory the value of the 4-forms is dual to the (quantized) value of internal fluxes. This
means that there is a quantized landscape of auxiliary fields in the effective field theory, and transitions between different vacua can in principle proceed through membrane nucleation.
(A  somewhat related approach has been also recently considered in \cite{Farakos:2016hly} involving in addition a nihilpotent multiplet and applying it to the
cancellation of the cosmological constant.)

Let us consider a toy  $N=1$ supergravity example with the required built-in  discrete shift  symmetries, consistent with 
having quantized 4-forms as auxiliary fields. Take a 2-field model with
\begin{equation}
  K \ =\ -2 \log (U+U^*)-3 \log (T+T^*) \ ,\   W\ = \ e_0 +\ i  h_0  U \ .
 \end{equation}
  With $U=u+ib$   the action will be invariant under the shift symmetry
 \beq
 b\ \rightarrow \ b\ +\ n  \ \ ;\ \  e_0\ \rightarrow \ e_0 \ +\  h_0 n    \ ,\ n\in {\bf Z}
 \eeq
 Consistency requires $e_0$ to be quantized in units of $h_0$. In string theory models these numbers are in general integers (see e.g. \cite{BOOK} and references therein),
 corresponding to quanta of internal fluxes, and we assume so in what follows.
 This is a no-scale model and the associated potential may be obtained in the standard way yielding
 \beq
 V\ = e^{K} G_{U{\bar U}}^{-1} |D_UW|^2 \ =\  {e^K} 2 | e_0-h_0 b|^2  \ .
 \eeq
 The potential  is shift invariant and has minima at $b=e_0/h_0$ in Minkowski,  and the rest of the fields are undetermined at this level. Supersymmetry is broken and the gravitino mass is given by
 \begin{equation}
 m_{3/2}^2=\dfrac{h_0^2}{2^5 t^3}.
 \end{equation}
 The standard $N=1$ auxiliary fields are given by 
 \beq
 F^U\ =\ e^{K/2} 2u (e_0\ -\ h_0 b)\ = \ 0 \ ;\  F^T\ =\ -ie^{K/2} 2h_0ut\ \not=\ 0 \ ,
 \eeq
 where $t=ReT$.
  With $h_0$ quantized we have a landscape of values for the gravitino mass (for fixed $u,t$).
 Note that the scalar potential of this system may be understood in terms of  a Minkowski 4-form with an action
 \beq
 {\cal L} _F\ =\ -    {e^{-K}}\ F_4^2 \ +\  2F_4(e_0\ -\  h_0 b) \ .
 \eeq
 Upon application of the equations 
 one obtains 
 \beq
 F_4 \ =\     e^{K}( e_0\ -\  h_0 b)  \ ,
 \eeq
 and the scalar potential above is recovered.
 The $N=1$ auxiliary field  for the $U$ field may be written in terms of this 4-form
 \beq
 F_U \ =\  2u e^{-K/2}\ F_4 \ .
 \eeq
   Still, since its vev is proportional to $h_0$, which is quantized, the gravitino mass and soft terms are also quantized.
 This is an example of a $N=1$ sugra model consistent   with a formulation in terms of 3-forms. Other examples 
 obtained from Type IIA and Type IIB orientifold vacua may be found in \cite{biv}.

We can consider now the addition of matter fields like e.g. a  MSSM Higgs sector $H_{u,d}$ and use the above toy model as a 'hidden sector'' for it.
If e.g. the Higgs fields had  minimal canonical kinetic terms
we will get for the Higgs mass  (see e.g. \cite{brignole}):
\beq
m^2_{H_u}=m^2_{H_d}\ = \ m^2_{3/2} \ =\    \frac{h_0^2M_s^4}{2^5M_p^2 t^3}.
\eeq
where we have  re-inserted the relevant Planck and string  mass factors which we were skipping up to now. 
 For values of $h_0 \simeq (10^{10}GeV)^2$, we will have qualitatively
\beq
m_H \ \simeq \   h_0\ \times \ 10^2 GeV  \left( \frac {M_s^2}{10^{20}GeV^2}\right) \ .
\eeq
Given that $h_0$ is quantized, the Higgs masses will scan in a landscape.
This  model is a 'toy''  since the rest of the scalar fields are undetermined, but that is inessential to the point we want to make, that there will be
 in general a landscape of Higgs masses if the auxiliary fields relevant of the hidden sector contain quantized 4-forms, as indicated by
 string theory.
 
 In general the full scalar potential in a fully realistic  MSSM depends on a variety of soft terms plus a $\mu_0$-term for the Higgs. At the end of the day, assuming 
 for simplicity flavour independence and universality, the mass of the  weak scale gauge bosons can be written as an expansion in terms of soft-terms
\beq
M_{Z^0}^2\,=\,c_1 M^2\,+\,c_2m^2\,+\,c_3|A|^2\,+\,c_4|\mu_0|^2 \,+\,c_5MA \,+\, c_6 \mu _0M \, +\, c_7 \mu_0 A \, \dots
\eeq
where $c_i=c_i(y_t,g_i)$ are coefficients depending only on the gauge and Yukawa couplings and including all the running between the UV scale and the 
EW scale. Here, in a standard notation,  $M$ is a universal gaugino mass, $m$ the soft scalar masses, $A$ is the trilinear soft coupling and
$\mu_0$ is the SUSY Higgs mass. In one such more complete setting 
all these soft terms ${\cal M}_{soft} ^{i}= m,M,A,B,\mu_0, ..$ will be quantized
\beq
{\cal M}_{soft}^{i} \ = n_i \frac {h_0^i }{M_p}\ ,\ n_i\in {\bf Z} \ ,\  
\eeq
where the $n_i$ of different soft terms need not be directly correlated, and the $h_0^i$ are of the same order. Thus we would have a grid of soft terms, with most of the points not giving appropriate 
EW symmetry breaking, but with some points consistent with correct EW breaking, with Higgs vevs consistent with anthropic considerations.

This built-in structure could perhaps explain  the little hierarchy problem. Indeed, it could be that soft terms could be above a few TeV, with squark and gluinos 
perhaps above LHC reach.  But 
for particular choices of the integers $n_i$, cancellations could take place allowing for correct EW symmetry breaking with an apparently fine-tuned choice of SUSY
parameters.

Let us comment about the connection between this SUSY landscape and the non-SUSY case considered in the previous section. 
In fact in the SUSY case, due to gravity mediation,  the coupling of the 4-forms to the Higgs mass is quadratic and Planck supresed, rather than linear. One indeed has couplings 
of the form $F_4^2|H|^2/M_p^2$, rather than $\eta F|H|^2$. One gets a mass  of order the EW scale for the axion in both cases if $\eta \simeq 10^{-8}$.
  There is also a difference in the kind of
landscape achieved. In the non-SUSY case there is a delicate cancellation between the UV mass $m^2$ of the Higgs and the contribution of the 4-forms $f^h,f^a$ as in eq.(\ref{masahiggs}).
If $m^2$ is very large one needs  $q_{a,h}\ll f_{a,h}$.   On the other hand in the above SUSY landscape  the value of the 4-forms (or auxiliary fields) is very large, naturally
(although not necessarily) of order $10^{20}$  GeV$^2$,   and the  fine-tuning is  naturally small, $\simeq F/M_p\simeq 100$ GeV because of gravity mediation.

\section{The SUSY breaking scale, the string scale  and the Weak Gravity Conjecture}

In this section we depart from the issue of the generation of a Higgs mass landscape and adopt a more general view, this time within string theory.
We would like to argue that in large classes of string compactifications (see e.g.\cite{BOOK}), the WGC suggests that there is a lower bound on the SUSY
breaking scale $m_{3/2}$ , depending on the string scale $M_s$. Roughly the lower bound is given by
\beq
m_{3/2} \ \gtrsim \  \frac {M_s^2}{M_p} \ .
\eeq
This bound has a number of loopholes and should only be considered as a common feature in certain classes of compactifications.
Still, since it would have important implications, we think it would be  worth studying how general it is. 

Let us summarize the underlying idea.
In the above $N=1$ SUSY examples the gravitino and soft-term masses scale like a quantized  parameter $h_0 $, which is an integer in units
of  some fundamental scale (i.e. $M_s^2$, with $M_s$ the string scale). In these models with quantized 4-forms 
there are membranes coupling to the 3-forms. 
The charge of these  membranes, of mass dimension two, will be proportional to $h_0$ or, reintroducing
the axion period $f$,  to $\mu f$, with $\mu$ the axion mass term. Now, as in the examples above, the WGC as applied to
3-forms and membranes give an upper bound on the tension  $T$ of the membranes coupling to a 3-form with charge $q$ \cite{imuv}
\beq
\frac {T}{M_p} \ \leq \ 2\pi q \ ,
\eeq
i.e., the strength of the 3-form coupling must be bigger than the gravitational coupling of the membrane. Applying these conditions to 
the axions $\phi_\alpha$  of some consistent string compactification one expects for all of them
\beq
\frac {T^\alpha}{f_\alpha M_p} \ \leq \ 2\pi  \mu_\alpha  
\label{primacota}
\eeq
as long as they couple to a massive  3-form.
This is interesting because it is telling us that all these axions cannot be arbitrarily light, since their mass corresponds to the
coupling of 3-forms to membranes, which cannot be small in order not to violate  the WGC. This should be preserved in any
consitent compactification.

In principle one can go case by case and  test in specific string compactifications whether the spectra of axion masses  respects the bounds (\ref{primacota}).
That may give relevant constraints on specific moduli fixing vacua and provide explicit tests of the WGC.
However one  can draw some general expectations from the given
structure. In particular, there are general classes of models in which axion  masses are directly related to the SUSY-breaking scale. 
Those are models in which the {\it Goldstino multiplet contains a monodromy axion}. In that case the mass of the axion is of the order 
of the gravitino mass and hence the bound applies not only to the axion but to the gravitino itself, i.e.
\beq
m_{3/2}\ \simeq \ m_\alpha \ \geq 	 \frac  {T_\alpha }{f_\alpha M_p}  \ .
\label{cotaya1}
\eeq
Models in which the Goldstino contains a monodromy axion include
Type IIA or Type IIB orientifolds with all moduli fixed by RR, NS and eventually additional geometric or 
non-geometric fluxes, see \cite{guarino} and references therein.  In these models SUSY is broken by the auxiliary fields of either Kahler, complex structure and/or complex dilaton.
Thus some  linear combination of axions will be SUSY partners of the Goldstino/gravitino, and the bound above would apply.
More generally, in typical string compactifications with broken SUSY and stabilized  moduli, either the Kahler, complex structure  or complex dilaton auxiliary fields tipically dominate SUSY breaking. In these cases  some linear combination of the axions in the moduli  will be a SUSY partner of the Goldstino/gravitino. So at least the mass of that particular linear combination   will be 
of order the gravitino mass, $m_a\simeq m_{3/2}$.

The bounds depend also on the membrane tensions and the periodicities.
 Concerning the axion periodicities $f_\alpha$, in string compactifications like these one typically has
$f_\alpha \simeq M_s$, although values as large as $M_p$ or slightly below $M_s$ are also possible, depending on volume factors.
Concerning the tensions of the RR membranes, they are in principle proportional to the volume wrapped by the higher dimensional  D-branes  or NS-branes yielding
membranes upon compactification.
One may argue that one can make the tensions arbitrarily small by making the cycles of the volumes arbitrarily small, which would make the bounds 
(\ref{primacota}) weaker.  However we would have to do that simultaneously with all the 3-forms and membranes, which sounds artificial.
Furthermore,  as emphasized already in  \cite{MR} , although the classical tensions can be vanishingly small, 
the effective tensions  are only slightly smaller than $M_s^3$. This is because the Weil-Peterson metric in e.g. a conifold cycle scales 
logarithmically with the blowing-up mode \cite{MR}.  In any event, let us evaluate the bounds by  setting the tensions $T_\alpha \simeq M_s^3$.
One gets
\beq
m_{3/2}\ \simeq  \ m_\alpha \ \geq 	 \frac  {T_\alpha }{f_\alpha M_p}\ \simeq \frac {M_{s}^2}{M_p}   \ .
\label{cotaya2}
\eeq
as advertised. We are also assuming here that $g_s\simeq 1$, as happens  in  semirealistic compactifications  in which one adjusts the gauge couplings
to the observed values. Thus we see that the scale of SUSY breaking cannot be arbitrarily low. That would imply the existence of interactions of some 3-form with membranes
with strength weaker than that of gravity.

The above bound, if true, would have important phenomenological implications. In this connection there are a couple of situations of particular phenomenological interest:

\begin{itemize}

\item {\it Intermediate scale SUSY breaking}. In this case SUSY is broken at $m_{3/2}\simeq 10^{12}$ GeV and $M_s\simeq 10^{15}$ GeV, consistent with the bound. 
The  spectrum below $m_{3/2}$ is that of the minimal SM.  This is interesting because it is known that, if one extrapolates the SM Higgs potential
corresponding to a 126 GeV Higgs at high energies, the potential develops an instability at around $10^{10}$ GeV  \cite{elias}.  If SUSY is restored above $10^{10}$ GeV 
such instability disappears.  This situation with an intermediate SUSY scale $M_{SS}$ has also been recently discussed both in the context of the 
observed SM Higgs mass  \cite{Ibanez:2012zg,Ibanez:2013gf} as well as in  MSSM Higgs inflation \cite{higgsotic}. In this situation no SUSY particles would be observed at LHC.

\item{\it SUSY at a TeV}. In this case one can have $m_{3/2}\simeq $ TeV with an intermediate string scale $M_s\simeq 10^{10}$, also consistent with the
bound. This is the case discussed in the previous section in the context of a MSSM landscape.
In this case SUSY particles could perhaps be observed at LHC but standard unification of coupling constants is lost.The case with an intermediate string scale has been considered  from different
considerations in the literature  (see e.g.  \cite{Burgess:1998px}). Note in particular that it was found in the context of Type IIB orientifolds with fluxes that SUSY breaking soft terms
are obtained for the MSSM scalars living on D7-branes which scale precisely in the same way, with \cite{Camara:2004jj}
\beq
m_{soft} \ \simeq \frac {f\ M_s^2}{M_p}
\eeq
and $f$ parametrizing the local fluxes on the brane positions. This soft terms would be consistent with the bound on SUSY breaking discussed above.

\end{itemize}

On the other hand a  big desert scenario with the GUT/String scale at $M_s\simeq 10^{16}$ GeV and low energy SUSY at a TeV, leading to succesful gauge coupling unification
would be inconsistent with such a bound. 

Let us close this section by noting that axions may get also a potential from instanton effects rather than directly from fluxes. This happens for example 
in Type IIB compactifications with standard NS and RR fluxes. The latter only induce monodromy to the complex structure and dilaton fields, but not to the 
axions in the Kahler multiplets. In the presence of gaugino condensation the role of the 3-forms is played  by the composite CS 3-form of the condensing gauge group, 
see \cite{Dvali1,Garcia-Valdecasas:2016voz}. In this case the associated membrane tension is of order $T\simeq \Lambda ^3$, with $\Lambda$ the condensate scale,
and the bound above constraint $\Lambda$ instead of the string scale. The origin of the composite  3-forms associated to non-gauge string instantons has been recently
worked out in \cite{Garcia-Valdecasas:2016voz}.

\section{Comments and conclusions}

In this paper we have studied how to generate a landscape of Higgs masses in order to address the gauge hierachy problem.
Although anthropic considerations based on the viability of complex nuclei constrain the Higgs vev to be close to the observed value,
we still need to have theories in which a landscape of Higgss masses, including viable ones, appear. This is what we tried to address in the present paper.

We put forward a general mechanism in which the landscape properties of an axion-3-form system is transmitted
to the Higgs sector of the SM or the MSSM. Indeed, the 4-form field strengths associated to 3-forms are assumed to be quantized,
as e.g. happens in string theory. On the other  hand there is an axion-like field which 1) gives a mass to the 3-form and 2) couples to
the Higgs field. Then the quantization properties of the axion/3-form system is transmitted to the Higgs sector via either a direct 
renormalizable coupling (as in a non-SUSY example discussed above) or  mediated by gravity, as in the SUSY examples discussed in the
previous section.

In the non-SUSY examples the mechanism suggests the existence of axion-like scalars with very weak couplings to the SM sector. 
Arguments based on the Weak Gravity Conjecture suggests masses for this { Hierarxion} not much below $10^{-3}$ eV, although the possible
range of values is very large.  In order to generate the landscape it is not needed that this axion couples directly to the QCD or photon field strengths,
as ordinary axions do.  On the other hand it can contribute to dark matter, although the chances to detect this axion with standard  techniques is model dependent.
One can contemplate the possibility of this axion to be identified with an ordinary PQ axion, but the fact that it couples to the 
Higgs sector makes difficult to achieve that goal, since its potential dominates over the standard instanton-induced potential. It would be interesting to
study different models in which different detection opportunities could be present.

In the SUSY examples, the axion/3-form system appears as part of the auxiliary fields involved in gravity mediation models. 
The presence of Minkowski 4-forms behaving as auxiliary fields of $N=1$ supergravity has been recently shown to be a general
property in string theory. In this case the philosophy is a bit different since the most natural situation is one in which the string scale
is identified with an intermediate scale  $M_s\simeq f\simeq 10^{10}$ GeV. The SUSY breaking soft terms are then of order ${\cal M}_{soft}\simeq F_4/M_p$
$\simeq M_s^2/M_p$
and scan in a landscape, with values of order the EW scale. There is a landscape of soft terms which could perhaps  provide a 
qualitative understanding of the SUSY fine-tuning implied by LHC results.

In both cases, SUSY and non-SUSY, the Weak Gravity Conjecture, as applied to 3-forms,  suggests that there is  a scale of new physics 
well below the Planck mass. Indeed, we saw that in the non-SUSY class of models  such an scale of order $\eta^{-1/3} 10^8$ GeV or below
should exist. In the SUSY case the string scale should typically be of order of the intermediate scale $10^{10}$ GeV or so, to generate a landscape.

More generally, one can argue that in large classes of string compactifications with fluxes the WGC suggests a lower bound
on the SUSY breaking scale with $m_{3/2}\gtrsim M_s^2/M_p$.  This applies in particular to models in which the 
Goldstino multiplet contains a monodromy axion, but it could be more general. 
Although, admittedly, there are a number of loopholes in such a bound, it would
be interesting to test it in specific compactifications.

Note that in here we have not addressed the problem of the cosmological constant. We are tacitaly assuming that there is a different mechanism, like
the Bousso-Polchinski (BP) mechanism  \cite{BP} which addresses this issue. Note that the mechanism discussed here is not of the BP type, in which delicate 
cancellations of a large (on the hundreds) multiplicity of 4-forms with large values, allows for the fine-tuning of the cosmological constant. 
One could think of the possibility of
addressing the issue of the c.c. in a way analogous to the mechanism discussed in the present paper.  However the scale of the cosmological constant 
is so small (of order $10^{-48}$ GeV$^4$) that  a threshold of new-physics associated to the required axion/3-form system should have been already
detected experimentally. We think on the other hand that a landscape for the EW sector appears more naturally in the context of
axion/3-form systems as here described.

\bigskip

\centerline{\bf \large Acknowledgments}

\bigskip

\noindent We thank  F. Marchesano, V. Mart\'{\i}n-Lozano, M. Montero, A. Uranga and I. Valenzuela for useful discussions. 
This work has been supported by the ERC Advanced Grant SPLE under contract ERC-2012-ADG-20120216-320421, by the grant FPA2012-32828 from the MINECO,  and the grant SEV-2012-0249 of the ``Centro de Excelencia Severo Ochoa" Programme.  The work of A. Herr\'aez is supported by a FPU fellowship from MINECO.


\end{document}